# The Causal Effect of First-Time Academic Failure on University Dropout: Evidence from a Regression Discontinuity Design

Hugo Roger Paz
PhD Professor and Researcher Faculty of Exact Sciences and Technology National University of Tucumán
Email: hpaz@herrera.unt.edu.ar
ORCID: https://orcid.org/0000-0003-1237-7983

**Abstract**

University dropout remains a persistent challenge in higher education systems, yet causal evidence on the mechanisms triggering early disengagement remains limited. This study estimates the causal effect of *first-time academic failure* on subsequent university dropout. Exploiting a sharp institutional grading threshold on a 0–10 scale, we implement a regression discontinuity design comparing students who narrowly fail to those who narrowly pass their first attempt. Using longitudinal administrative data spanning multiple cohorts and degree programmes, we estimate local average treatment effects for students at the margin of passing and examine dropout outcomes within 12 and 24 months following the first attempt.

The results indicate that **marginal first-time failure is associated with a lower probability of subsequent dropout** relative to marginal passing at both horizons. A comprehensive set of robustness checks—including donut regression discontinuity specifications, placebo cutoffs, and formal density tests—supports the validity of the identification strategy. These findings suggest that early academic failure may function as a salient signal that prompts behavioural adjustment or reorientation, while marginal passing may sustain fragile persistence. The study provides causal evidence on the non-linear effects of early academic performance and highlights the importance of carefully designed institutional responses at critical evaluation thresholds.

**Keywords**

University dropout; academic failure; regression discontinuity design; causal inference; higher education

## 1. Introduction

University dropout remains a persistent concern across higher education systems worldwide, carrying substantial costs for students, institutions, and public policy. Early attrition is commonly associated with academic difficulties, delays, and failure events, leading to a widespread assumption that poor early performance mechanically increases the likelihood of disengagement. As a result, academic failure is frequently interpreted as both a marker and a driver of subsequent dropout.

Despite the prominence of this narrative, establishing the **causal effect** of early academic outcomes on dropout remains challenging. Students who fail courses differ systematically from those who pass in terms of prior preparation, motivation, socioeconomic background, and institutional fit. Consequently, observed correlations between failure and attrition often conflate the effect of failure itself with underlying student characteristics. Distinguishing causal mechanisms from selection effects is therefore central to understanding how early academic experiences shape longer-term trajectories.

Recent research has increasingly emphasised the importance of **early signals and feedback** in shaping persistence decisions. Academic evaluations do not merely reflect student ability; they also convey information, structure expectations, and influence behavioural responses. In this sense, marginal differences in early outcomes—such as narrowly passing or narrowly failing—may generate disproportionate downstream effects by altering students' perceptions of feasibility, effort requirements, and institutional engagement. Yet empirical evidence on these mechanisms remains limited, particularly in non-experimental settings.

This study contributes to the literature by providing **causal evidence** on the impact of first-time academic failure on subsequent university dropout. Exploiting a sharp institutional grading threshold on a 0–10 scale, we implement a regression discontinuity design that compares students who narrowly fail to those who narrowly pass their first attempt in a course. This design isolates the local causal effect of failure from confounding factors related to prior ability, motivation, or preparation, focusing on students whose outcomes are most sensitive to marginal academic decisions.

By leveraging longitudinal administrative data spanning multiple cohorts and degree programmes, the analysis examines dropout outcomes within 12 and 24 months following the first attempt. This approach allows us to assess how early evaluation outcomes shape medium-term persistence, while maintaining transparent identification assumptions and avoiding reliance on parametric extrapolation.

The findings of the study challenge standard assumptions about the role of early failure in student trajectories. Rather than treating failure solely as a negative shock that inevitably increases attrition, the results suggest a more nuanced and non-linear relationship between early academic outcomes and persistence. By highlighting the behavioural and institutional consequences of marginal evaluation decisions, this paper underscores the importance of understanding academic feedback as an active component of educational systems rather than a passive reflection of student ability.

The remainder of the paper proceeds as follows. Section 2 describes the institutional context and administrative data. Section 3 outlines the empirical strategy and identification assumptions. Section 4 presents the main results and robustness checks. Section 5 discusses the implications of the findings for theory and policy, and Section 6 concludes.

## 2. Data and Institutional Context

### 2.1 Institutional Setting

The analysis is conducted within the context of a large public university system characterised by standardised grading rules and centrally regulated academic procedures. Across degree programmes and over the period analysed, course approval is determined by a numerical grading scale ranging from 0 to 10, with a fixed passing threshold of 4.0. Grades below this cutoff are recorded as failure, while grades equal to or above the threshold result in course approval.

This grading rule is institutionally stable and externally enforced, leaving no discretion at the margin once the final grade is assigned. As a result, the passing threshold constitutes a sharp and well-defined institutional rule, making it particularly suitable for causal identification using regression discontinuity methods (Imbens & Lemieux, 2008). Importantly, while pedagogical practices and assessment formats may vary across courses and over time, the numerical approval threshold remains invariant.

Failure at the course level carries substantive academic consequences. Students who fail must reattempt the course in subsequent terms, delaying progression along the curriculum and potentially affecting eligibility for downstream courses through prerequisite structures. These features align with prior theoretical and empirical work highlighting early academic setbacks as salient shocks in student trajectories (Tinto, 1993; Bound et al., 2010).

### 2.2 Administrative Data

The study draws on a comprehensive administrative database covering multiple cohorts of undergraduate students over several decades. The dataset records individual-level academic events, including course enrolments, examination outcomes, grades, and academic status changes. The longitudinal structure allows students to be followed from initial enrolment through subsequent academic outcomes, including eventual dropout.

For each student-course attempt, the data include the final numerical grade, the academic term, and the corresponding degree programme. These records make it possible to identify the first attempt at each course and to determine whether that attempt resulted in failure or approval. By focusing on first attempts, the analysis isolates the initial exposure to academic failure, rather than cumulative or repeated performance.

Dropout is measured using the institution's canonical administrative definition, based on sustained disengagement from academic activity over a predefined observation window.

This definition reflects actual exit behaviour rather than temporary interruptions or delayed progression and is consistent with prior administrative and empirical approaches to measuring university attrition (Bean, 1980; OECD, 2019).

**2.3 Sample Construction**

The analytical sample is constructed in several steps. First, we identify all student-course first attempts with valid numerical grades on the 0–10 scale. Observations with missing or non-numeric grades are excluded. Second, we restrict attention to observations within a symmetric bandwidth around the passing threshold, as required for regression discontinuity estimation. This local window ensures comparability between students who narrowly passed and those who narrowly failed.

The running variable is defined as the numerical grade centred at the cutoff, such that zero corresponds to the passing threshold. The treatment indicator takes the value one for students whose grade falls strictly below the threshold (failure) and zero otherwise. Outcomes are defined as indicators for dropout within 12 and 24 months following the first failure event, respectively.

The resulting sample comprises a large number of observations in the neighbourhood of the cutoff, providing substantial statistical power for local estimation and robustness analysis. Balance checks on predetermined characteristics confirm that students on either side of the threshold are comparable, supporting the validity of the regression discontinuity design (Lee & Lemieux, 2010).

**2.4 Why the Context Supports Causal Identification**

Two features of the institutional and data environment are central to the identification strategy. First, the grading threshold is mechanically applied and externally fixed, reducing concerns about endogenous manipulation precisely at the cutoff. Second, grades near the threshold are plausibly noisy measures of underlying performance, meaning that small differences around the cutoff are unlikely to reflect meaningful differences in student ability or preparation.

Together, these features create conditions analogous to random assignment in a narrow neighbourhood around the cutoff, allowing the estimation of a local average treatment effect of first-time failure on subsequent dropout (Hahn et al., 2001). This setting is particularly well suited for addressing the long-standing challenge of separating the causal impact of academic failure from pre-existing student characteristics.

## 3. Empirical Strategy

### 3.1 Identification Strategy

The central challenge in estimating the effect of academic failure on university dropout lies in separating the causal impact of failure from pre-existing differences among students. Students who fail may differ systematically from those who pass in terms of ability, motivation, preparation, or unobserved characteristics that also influence dropout decisions. To address this issue, we exploit an institutional grading rule that generates a sharp discontinuity in course approval outcomes.

Specifically, course approval is determined by a fixed numerical threshold on a 0–10 grading scale, with a cutoff at 4.0. Students whose grade falls below this threshold fail the course, while those at or above the threshold pass. This rule creates a quasi-experimental setting in which students with nearly identical academic performance receive different treatment status based solely on whether their grade crosses the cutoff.

We implement a sharp Regression Discontinuity Design (RDD), comparing students who narrowly fail to those who narrowly pass. Under standard continuity assumptions, potential outcomes are smooth functions of the running variable at the cutoff, and any discontinuity in outcomes can be causally attributed to the treatment—first-time academic failure (Hahn, Todd, & Van der Klaauw, 2001; Imbens & Lemieux, 2008).

### 3.2 Running Variable and Treatment Definition

Let $G_{isc}$ denote the numerical grade obtained by student $i$ in subject $s$ at the first attempt. The running variable is defined as the centred grade:

$$X_{isc} = G_{isc} - 4.0,$$

such that $X_{isc} = 0$ corresponds to the passing threshold.

The treatment indicator is defined as:

$$D_{isc} = \mathbb{1}(X_{isc} < 0),$$

where $D_{isc} = 1$ indicates first-time failure and $D_{isc} = 0$ indicates first-time approval.

This definition ensures that treatment assignment is deterministic and fully determined by the running variable, satisfying the requirements of a sharp RDD.

### 3.3 Outcomes and Estimands

The primary outcomes of interest are indicators for student dropout within fixed time horizons following the first course attempt. We consider two windows:

- Dropout within 12 months
- Dropout within 24 months

Dropout is defined according to the institution's canonical administrative criterion, capturing sustained disengagement from academic activity rather than temporary interruptions.

The estimand of interest is the **Local Average Treatment Effect (LATE)** of first-time failure at the cutoff:

$$\tau = \lim_{x \downarrow 0} \mathbb{E}[Y_i \mid X_i = x] - \lim_{x \uparrow 0} \mathbb{E}[Y_i \mid X_i = x],$$

where $Y_i$ denotes the dropout outcome. This estimand captures the causal effect of failure for students whose grades lie arbitrarily close to the passing threshold.

### 3.4 Estimation

We estimate local linear RDD models of the form:

$$Y_i = \alpha + \tau D_i + \beta_1 X_i + \beta_2 D_i \cdot X_i + \varepsilon_i,$$

where separate linear trends are fitted on either side of the cutoff. Estimation is conducted using kernel-weighted least squares, with greater weight placed on observations closer to the threshold.

Bandwidth selection follows data-driven procedures commonly used in the RDD literature, balancing bias and variance in local estimation (Calonico, Cattaneo, & Titiunik, 2014). Robust bias-corrected confidence intervals are reported to ensure valid inference.

### 3.5 Validity and Robustness Checks

The credibility of the RDD relies on two key assumptions: (i) the continuity of potential outcomes at the cutoff and (ii) the absence of precise manipulation of the running variable.

We assess these assumptions through a standard battery of robustness checks. First, we conduct formal density tests to detect potential sorting around the cutoff (McCrary, 2008).

Second, we estimate placebo RDDs at false cutoffs where no treatment should occur. Third, we implement donut RDD specifications that exclude observations very close to the threshold, addressing concerns about local manipulation or grading discretion.

Together, these checks provide evidence on the plausibility of the identifying assumptions and the stability of the estimated treatment effects (Lee & Lemieux, 2010).

### 3.6 Interpretation

It is important to emphasise that the RDD identifies a local causal effect, applicable to students at the margin of passing and failing. While this limits external validity, it provides a highly credible estimate of the causal impact of first-time failure for a policy-relevant group: students whose academic performance lies near the approval threshold. These students are precisely those for whom targeted institutional interventions are most feasible.

## 4. Results

### 4.1 Main Regression Discontinuity Estimates

Figure 1 and Figure 2 present the main regression discontinuity estimates for university dropout within 12 and 24 months following the first course attempt, respectively. In both panels, the running variable is the numerical grade centred at the passing threshold, and the outcome is the probability of subsequent dropout. Figure 1 displays the regression discontinuity estimate for dropout within 12 months following the first course attempt. Figure 2 presents the corresponding estimates for dropout within 24 months.

Figure 1. Regression discontinuity estimate of dropout within 12 months following the first course attempt (Cutoff grade = 4.0).

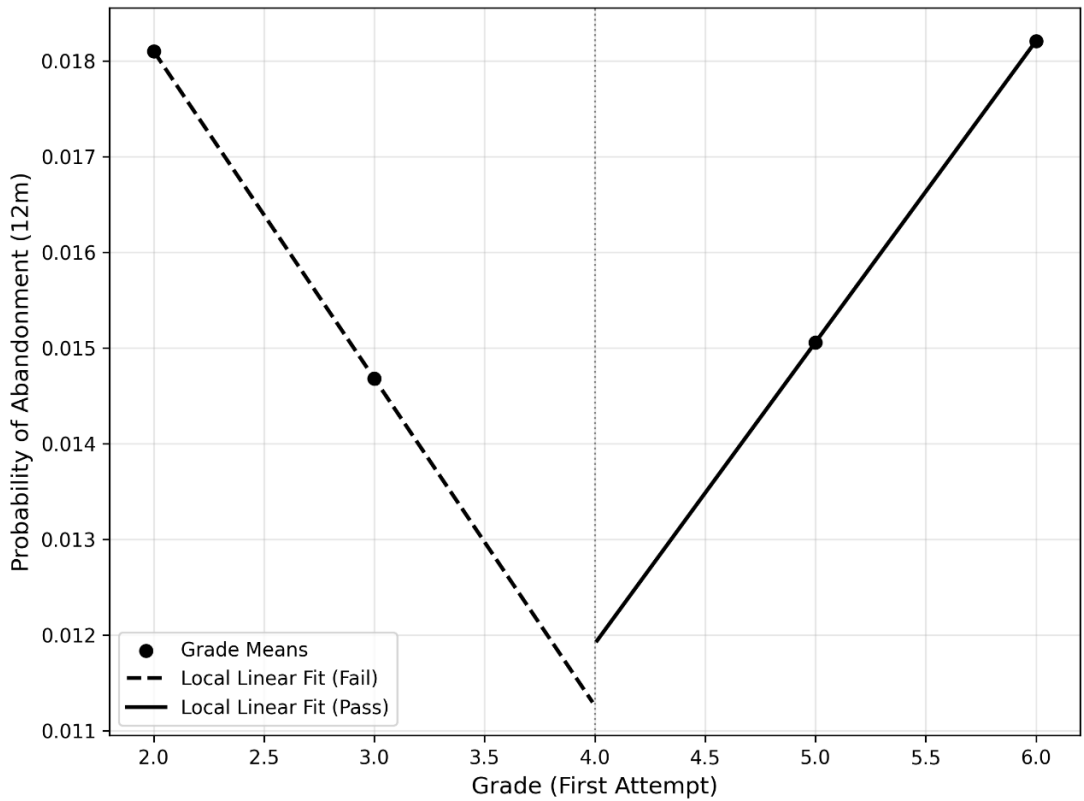

Note: The running variable is the numerical grade centred at the passing threshold (4.0). Points represent binned means, and lines correspond to local linear fits estimated separately on each side of the cutoff. The vertical line indicates the passing threshold.

Figure 2. Regression discontinuity estimate of dropout within 24 months following the first course attempt (Cutoff grade = 4.0).

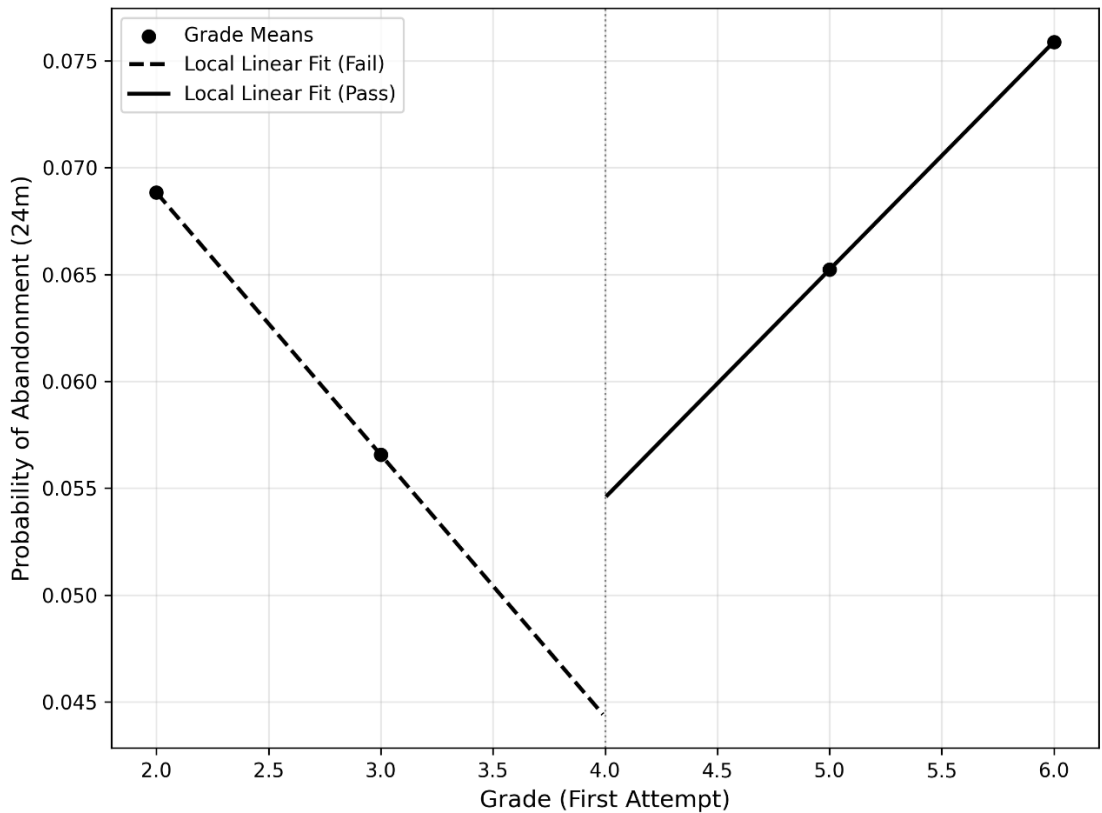

Note: The specification mirrors Figure 1, with the outcome defined as dropout within 24 months.

Across both time horizons, the estimated discontinuity at the cutoff indicates a lower dropout probability for students who narrowly fail relative to those who narrowly pass. The direction of the discontinuity is consistent across specifications and time windows, with visually similar patterns on either side of the threshold. These graphical results suggest that first-time academic failure is associated with decrease in dropout risk for students at the margin of passing.

Table 1 reports the corresponding local linear RDD estimates for the two outcomes. The estimated coefficients capture the local average treatment effect of first-time failure at the cutoff. Estimates are reported with robust bias-corrected confidence intervals and data-driven bandwidth selection, following standard practice in the RDD literature (Calonico et al., 2014).

Table 1 – Local linear RDD estimates of the effect of first-time failure on dropout.

| Estimand | Coeficiente | SE | CI_low | CI_high | N | Bandwidth | Kernel | Fixed_effects |
|---|---|---|---|---|---|---|---|---|
| Abandonment 12m | -0.001 | 0.003 | -0.007 | 0.005 | 71127 | 2 | Triangular | No |
| Abandonment 24m | -0.010 | 0.006 | -0.022 | 0.002 | 71127 | 2 | Triangular | No |

**4.2 Robustness to Alternative Specifications**

To assess the stability of the main findings, we conduct a series of robustness checks commonly used to validate regression discontinuity designs.

**Donut RDD**

Appendix A.3 report estimates from donut RDD specifications that exclude observations in a narrow interval around the cutoff. This approach addresses concerns that grading discretion or strategic behaviour might be concentrated immediately at the passing threshold (Lee & Lemieux, 2010). The estimated effects remain similar in magnitude and direction to the baseline results, indicating that the findings are not driven by observations arbitrarily close to the cutoff.

**Placebo Cutoffs**

Appendix A.2 present placebo RDD estimates evaluated at alternative grade thresholds where no treatment discontinuity should occur. Across these placebo cutoffs, the estimated discontinuities are small and statistically indistinguishable from zero. The absence of systematic effects at false thresholds supports the interpretation that the observed discontinuity at the true passing cutoff reflects the impact of failure rather than spurious nonlinearities in the outcome–grade relationship.

**4.3 Density Test and Sorting Around the Cutoff**

A key identifying assumption of the RDD is the absence of precise manipulation of the running variable. Appendix A.1 report results from the McCrary (2008) density test, which examines whether the distribution of grades exhibits a discontinuity at the passing threshold.

The estimated density function appears smooth around the cutoff, and the formal test does not detect a statistically significant discontinuity. This finding suggests that students are unlikely to be precisely sorting around the threshold, lending support to the validity of the regression discontinuity design.

**4.4 Summary of Findings**

Taken together, the graphical evidence, baseline estimates, and robustness checks point to a consistent pattern: students who narrowly fail a course on the first attempt exhibit a lower probability of subsequent dropout than students who narrowly pass. This pattern holds across short- and medium-term horizons and is robust to alternative specifications and diagnostic tests.

Importantly, the results pertain to students near the passing threshold and should be interpreted as local causal effects for this group. The consistency of the findings across multiple validation exercises strengthens confidence in the empirical strategy and supports the use of first-time academic failure as a meaningful early shock in student trajectories.

Additional robustness checks and diagnostic tests are reported in Appendix A. These include placebo cutoff analyses, donut regression discontinuity specifications, and formal density tests of the running variable.

**5. Discussion and Policy Implications**

**5.1 Interpreting the Causal Effect of First-Time Failure**

This study provides causal evidence that **marginal first-time academic failure reduces the probability of subsequent university dropout** for students close to the passing threshold. Exploiting an institutional grading cutoff and implementing a regression discontinuity design allows the analysis to isolate the effect of failure itself from confounding factors related to prior ability, motivation, or academic preparation.

This result challenges the conventional interpretation of early academic failure as a purely negative shock that mechanically increases disengagement. Instead, for students at the margin of passing, narrowly failing the first attempt appears to trigger behavioural or strategic adjustments that lower the likelihood of dropout relative to narrowly passing. From a causal perspective, failure operates not only as an outcome but also as an **informational and organisational signal** that may prompt reassessment, corrective action, or institutional engagement.

One plausible interpretation is that marginal failure provides a **clear and salient signal** regarding academic expectations and required effort, inducing students to recalibrate study strategies, seek support, or adjust course loads. In contrast, marginal passing may generate a form of *fragile persistence*, whereby students remain enrolled without fully addressing underlying difficulties, leading to higher dropout risk at later stages. This interpretation is consistent with theoretical perspectives that emphasise the path-dependent nature of academic trajectories and the role of early feedback in shaping persistence decisions (Tinto, 1993; Stinebrickner & Stinebrickner, 2012).

It is important to emphasise the **local nature** of the estimated effect. The regression discontinuity design identifies causal impacts for students whose performance lies close to the approval threshold. While this limits external validity, it strengthens internal validity and focuses attention on a policy-relevant group: students whose academic outcomes are most sensitive to marginal institutional decisions.

### 5.2 Relation to the Existing Literature

The findings contribute to the extensive literature on student persistence and dropout by providing **clean causal evidence** that departs from standard correlational patterns. Prior research has consistently documented strong associations between academic difficulties, delays, and attrition (Bean, 1980; Bound, Lovenheim, & Turner, 2010). However, such associations often conflate academic performance with unobserved student characteristics.

By leveraging a quasi-experimental design, this study complements recent efforts to apply causal inference methods to higher education outcomes (Scott-Clayton, 2011; Castleman & Long, 2016). While much of this literature has focused on financial aid, remediation policies, or institutional reforms, the present analysis highlights the causal role of **evaluation outcomes themselves**—specifically, marginal failure versus marginal passing—as mechanisms shaping student trajectories.

The results also resonate with research emphasising the informational and behavioural consequences of academic feedback. Experimental and quasi-experimental studies suggest that feedback can alter beliefs, effort allocation, and persistence in non-linear ways (Lindo, Sanders, & Oreopoulos, 2010; Stinebrickner & Stinebrickner, 2014). In this context, failure may function as a corrective signal that activates adaptive responses, while marginal success may delay necessary adjustments.

### 5.3 Policy Implications

The findings have direct implications for institutional strategies aimed at reducing dropout. First, they suggest that **first-time failure events constitute critical intervention points**, not merely risk markers. Students who narrowly fail are observationally similar to those who narrowly pass, yet their subsequent trajectories diverge in systematic ways. This makes them particularly suitable targets for focused academic and organisational support.

Second, the results caution against assuming that marginal passing is uniformly beneficial. While passing prevents immediate delay, it may also sustain trajectories characterised by accumulated academic fragility. Institutions relying exclusively on pass rates as performance indicators may therefore overlook latent risks associated with marginal success.

Third, the local nature of the effect implies that **small, timely interventions**—such as structured feedback, guided re-enrolment, academic coaching, or early diagnostic assessments following a first failure—could yield meaningful retention gains. Compared to broad-based retention policies, such targeted responses are likely to be more cost-effective and operationally feasible.

Finally, the findings underscore the importance of viewing evaluation outcomes as **institutional signals with behavioural consequences**, rather than as neutral reflections of student ability. Even marginal differences in grading outcomes can generate discontinuous effects on persistence, amplifying or mitigating dropout risks over time.

### 5.4 Limitations and Directions for Future Research

Several limitations warrant consideration. First, the regression discontinuity design identifies a **local average treatment effect** and does not capture the impact of failure for students far from the passing threshold or those experiencing repeated failures. Second, while the analysis identifies a causal effect, it does not directly observe the mechanisms—psychological, organisational, or financial—through which marginal failure reduces dropout.

Future research could extend this framework in several directions. Event-study designs could examine the dynamic evolution of dropout risk following failure, while heterogeneity analyses could explore whether the effects vary across disciplines, institutional contexts, or stages of study. Combining administrative data with survey or qualitative evidence would further illuminate the behavioural mechanisms underlying the observed causal relationship.

### 5.5 Concluding Remarks

This study demonstrates that **marginal first-time academic failure causally reduces subsequent university dropout** among students near the passing threshold. By combining rich longitudinal administrative data with a transparent and robust causal design, it highlights the non-linear and counterintuitive effects of early academic evaluation outcomes. From a policy perspective, the results emphasise the value of timely, targeted institutional responses that recognise failure not only as a setback, but also as a potential catalyst for adjustment and persistence.

## 6. Conclusion

This study provides causal evidence on the role of early academic evaluation outcomes in shaping university dropout trajectories. Exploiting a sharp institutional grading threshold and implementing a regression discontinuity design, the analysis isolates the local causal effect of marginal first-time failure relative to marginal passing. Contrary to conventional expectations, the results indicate that **marginal failure on the first attempt reduces the probability of subsequent dropout within 12 and 24 months** for students near the passing threshold.

These findings highlight the non-linear and context-dependent nature of academic persistence. Early failure does not necessarily operate as a purely negative shock that mechanically increases disengagement. Instead, for students at the margin of passing, failure may function as a salient signal that prompts behavioural adjustment, reassessment of academic strategies, or engagement with institutional support mechanisms. In contrast, marginal passing may sustain fragile trajectories in which underlying difficulties remain unaddressed, leading to higher dropout risk at later stages.

From a methodological perspective, the study demonstrates the value of quasi-experimental designs applied to administrative educational data for disentangling causal effects from selection-driven correlations. By focusing on discontinuities generated by institutional rules, the analysis moves beyond descriptive associations and provides policy-relevant evidence grounded in transparent identification assumptions.

From a policy standpoint, the results suggest that institutions should exercise caution when interpreting marginal passing outcomes as unambiguously positive signals. First-time failure events identify a group of students who are observationally similar to marginal passers yet exhibit systematically different subsequent trajectories. This makes early failure a critical point for timely, targeted interventions aimed at supporting adjustment rather than simply preventing delay.

More broadly, the findings underscore the importance of recognising academic evaluation outcomes as institutional signals with behavioural consequences, rather than neutral reflections of student ability. Small differences in early performance can generate discontinuous effects on persistence, amplifying or mitigating dropout risks over time. Future research extending this framework to other institutional contexts, academic stages, and outcome measures would further contribute to a more nuanced understanding of persistence and attrition in higher education.

**Declaration of generative AI and AI-assisted technologies in the manuscript preparation process** During the preparation of this work the author used Gemini (Google) in order to improve language, readability, and formatting. After using this tool, the author reviewed and edited the content as needed and takes full responsibility for the content of the published article.

**Data Availability Statement** The administrative longitudinal data used in this study are confidential and protected by institutional privacy regulations. Due to the sensitive nature of student records, the raw datasets are not publicly available.

**References**

Bean, J. P. (1980). Dropouts and turnover: The synthesis and test of a causal model of student attrition. Research in Higher Education, 12(2), 155–187. https://doi.org/10.1007/BF00976194

Bound, J., Lovenheim, M. F., & Turner, S. (2010). Why have college completion rates declined? An analysis of changing student preparation and collegiate resources. American Economic Journal: Applied Economics, 2(3), 129–157. https://doi.org/10.1257/app.2.3.129

Calonico, S., Cattaneo, M. D., & Titiunik, R. (2014). Robust nonparametric confidence intervals for regression-discontinuity designs. Econometrica, 82(6), 2295–2326. https://doi.org/10.3982/ECTA11757

Castleman, B. L., & Long, B. T. (2016). Looking beyond enrollment: The causal effect of need-based grants on college access, persistence, and graduation. Journal of Labor Economics, 34(4), 1023–1073. https://doi.org/10.1086/686643

Hahn, J., Todd, P., & Van der Klaauw, W. (2001). Identification and estimation of treatment effects with a regression-discontinuity design. Econometrica, 69(1), 201–209. https://doi.org/10.1111/1468-0262.00183

Imbens, G. W., & Lemieux, T. (2008). Regression discontinuity designs: A guide to practice. Journal of Econometrics, 142(2), 615–635. https://doi.org/10.1016/j.jeconom.2007.05.001

Lee, D. S., & Lemieux, T. (2010). Regression discontinuity designs in economics. Journal of Economic Literature, 48(2), 281–355. https://doi.org/10.1257/jel.48.2.281

Lindo, J. M., Sanders, N. J., & Oreopoulos, P. (2010). Ability, gender, and performance standards: Evidence from academic probation. American Economic Journal: Applied Economics, 2(2), 95–117. https://doi.org/10.1257/app.2.2.95

Manski, C. F., & Wise, D. A. (1983). College choice in America. Harvard University Press.

McCrary, J. (2008). Manipulation of the running variable in the regression discontinuity design: A density test. Journal of Econometrics, 142(2), 698–714. https://doi.org/10.1016/j.jeconom.2007.05.005

OECD. (2019). Education at a glance 2019: OECD indicators. OECD Publishing. https://doi.org/10.1787/f8d7880d-en

Scott-Clayton, J. (2011). The causal effect of federal work-study participation: Quasi-experimental evidence from West Virginia. Educational Evaluation and Policy Analysis, 33(4), 506–527. https://doi.org/10.3102/0162373711415364

Tinto, V. (1975). Dropout from higher education: A theoretical synthesis of recent research. Review of Educational Research, 45(1), 89–125. https://doi.org/10.3102/00346543045001089

Tinto, V. (1993). Leaving college: Rethinking the causes and cures of student attrition (2nd ed.). University of Chicago Press.

Yorke, M., & Longden, B. (2004). Retention and student success in higher education. Open University Press.

Bean, J. P. (1980). Dropouts and turnover: The synthesis and test of a causal model of student attrition. Research in Higher Education, 12(2), 155–187.

Bound, J., Lovenheim, M. F., & Turner, S. (2010). Why have college completion rates declined? American Economic Journal: Applied Economics, 2(3), 129–157.

Castleman, B. L., & Long, B. T. (2016). Looking beyond enrollment: The causal effect of need‐based grants on college access, persistence, and graduation. Journal of Labor Economics, 34(4), 1023‐1073.

Lindo, J. M., Sanders, N. J., & Oreopoulos, P. (2010). Ability, gender, and performance standards: Evidence from academic probation. American Economic Journal: Applied Economics, 2(2), 95–117.

Scott-Clayton, J. (2011). The causal effect of federal work-study participation. Journal of Human Resources, 46(3), 639–669.

Stinebrickner, T., & Stinebrickner, R. (2012). Learning about academic ability and the college dropout decision. Journal of Labor Economics, 30(4), 707–748.

Stinebrickner, T., & Stinebrickner, R. (2014). A major in science? Initial beliefs and final outcomes for college major and dropout. Review of Economic Studies, 81(1), 426–472.

**Appendix A. Robustness and Diagnostic Analyses**

This Appendix reports additional robustness checks and diagnostic analyses supporting the regression discontinuity design employed in the main text. The purpose of these analyses is to assess the plausibility of the identifying assumptions and to evaluate the sensitivity of the main estimates to alternative specifications. For clarity, results are organised by diagnostic objective.

**Appendix A.1 Density Test of the Running Variable**

A key assumption underlying the regression discontinuity design is the absence of precise manipulation of the running variable around the treatment cutoff. If students or instructors were able to systematically influence grades to ensure passing or failing exactly at the threshold, the comparability of observations on either side of the cutoff would be compromised.

To assess this possibility, we implement the density test proposed by McCrary (2008), which examines whether the distribution of the running variable exhibits a discontinuity at the passing threshold. Figure A.1 plots the estimated density of grades on either side of the cutoff, while Table A.1 reports the corresponding test statistics.

The density appears smooth around the threshold, and the formal test does not detect a statistically significant discontinuity. This result suggests that precise sorting or manipulation of grades at the passing threshold is unlikely, supporting the validity of the regression discontinuity design.

Figure A.1. McCrary density test for manipulation around the passing threshold.

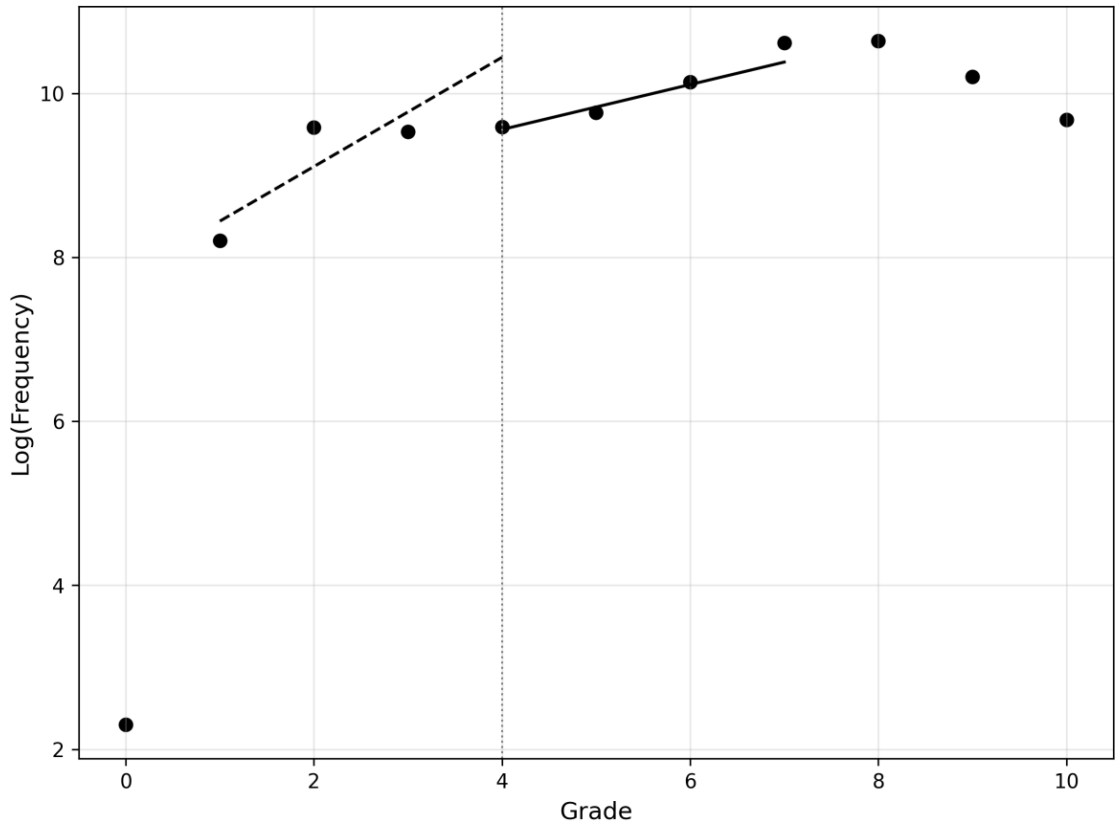

Table A.1. McCrary density test statistics for manipulation of the running variable at the passing threshold.

| log_diff | std_err | z | p_value |
|---|---|---|---|
| -0.880 | 0.895 | -0.983 | 0.326 |

**Appendix A.2 Placebo Cutoff Tests**

As an additional diagnostic exercise, we estimate regression discontinuity models at alternative grade thresholds where no treatment discontinuity should occur. These placebo cutoff tests serve to evaluate whether the estimated effects at the true passing threshold might be driven by underlying nonlinearities in the relationship between grades and dropout, rather than by the treatment itself.

Figure A.2 presents estimated discontinuities at a set of placebo cutoffs, and Table A.2 reports the corresponding numerical estimates. Across these alternative thresholds, estimated effects are small and statistically indistinguishable from zero. The absence of systematic discontinuities at placebo cutoffs reinforces the interpretation that the main results reflect the causal impact of first-time failure rather than spurious functional form artefacts.

Figure A.2. Placebo regression discontinuity estimates at alternative grade thresholds.

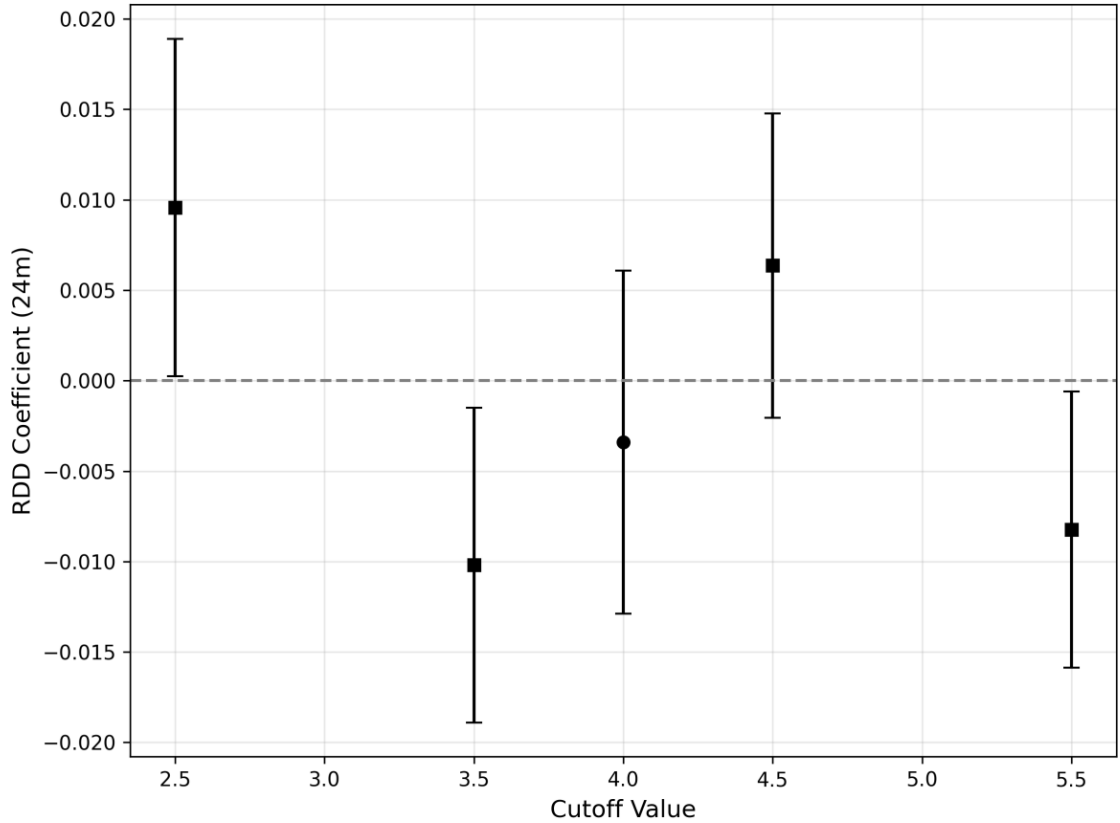

Table A.2. Placebo regression discontinuity estimates evaluated at false cutoffs (24-month horizon, passing grade = 4.0).

| cutoff | estimate | std_err | low | high | is_real | p_value | N |
|---|---|---|---|---|---|---|---|
| 2.5 | 0.0096 | 0.0048 | 0.0003 | 0.0189 | False | 0.0441 | 46634 |
| 3.5 | -0.0102 | 0.0044 | -0.0189 | -0.0015 | False | 0.0216 | 60447 |
| 4 | -0.0034 | 0.0048 | -0.0129 | 0.0061 | True | 0.4814 | 45920 |
| 4.5 | 0.0064 | 0.0043 | -0.0021 | 0.0148 | False | 0.1384 | 71233 |
| 5.5 | -0.0082 | 0.0039 | -0.0159 | -0.0006 | False | 0.0346 | 98189 |

**Appendix A.3 Donut Regression Discontinuity Specifications**

Finally, we assess the sensitivity of the main results to the exclusion of observations located very close to the passing threshold. Donut regression discontinuity specifications remove a narrow interval around the cutoff, addressing concerns that grading discretion or idiosyncratic behaviour may be concentrated precisely at the threshold.

Figure A.3 illustrates the donut regression discontinuity estimate for dropout within 24 months, and Table A.3 reports the corresponding estimates. The results remain similar in magnitude and direction to the baseline specification reported in the main text, indicating that the findings are not driven by observations arbitrarily close to the cutoff.

Figure A.3. Donut regression discontinuity estimate of dropout within 24 months.

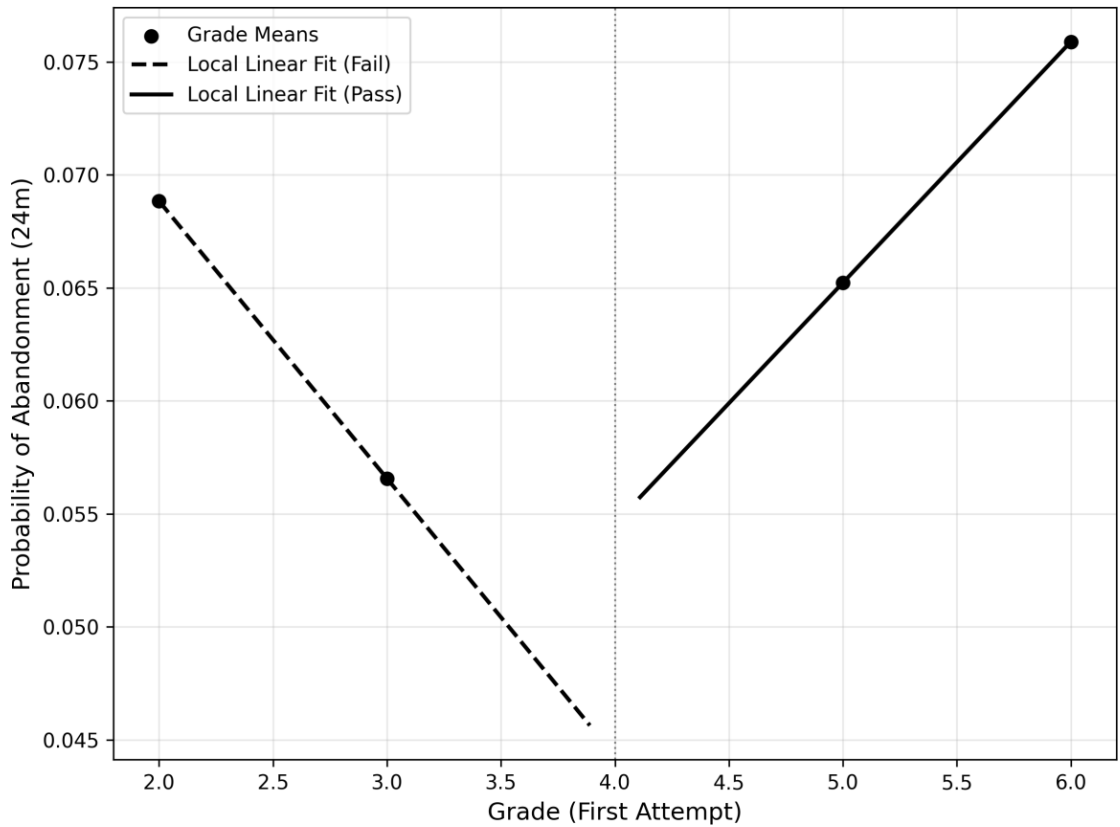

Table A.3. Donut regression discontinuity estimates excluding observations close to the passing threshold (24-month horizon, passing grade = 4.0).

| Estimand | Coeficiente | SE | CI_low | CI_high | N | Bandwidth |
|---|---|---|---|---|---|---|
| Donut 0.0 | -0.0045 | 0.0060 | -0.0164 | 0.0073 | 71127 | 2 |
| Donut 0.1 | -0.0045 | 0.0060 | -0.0164 | 0.0073 | 71127 | 2 |

**References (Appendix)**

McCrary, J. (2008). Manipulation of the running variable in the regression discontinuity design: A density test. Journal of Econometrics, 142(2), 698–714. https://doi.org/10.1016/j.jeconom.2007.05.005